\documentclass[12pt,preprint]{aastex}
\usepackage{natbib}
\usepackage{graphicx}
\shorttitle{The exotic BH-NS binaries in our Galaxy}
\begin{document}
\title{The exotic black hole-neutron star binaries in our Galaxy}
\author{Man Ho Chan}
\affil{Department of Science and Environmental Studies, The Education University of Hong Kong, Hong Kong, China}
\email{chanmh@eduhk.hk}

\begin{abstract}
It has been suggested that there are $\sim 10^5$ black hole-neutron star (BH-NS) binaries in our Galaxy. However, despite the effort of intensive radio search for decades, none of these binaries has been found to date. These binaries are regarded as a holy grail of astronomy because they can greatly improve our understanding about relativistic systems of compact objects and fundamental physics. In this article, we propose the existence of exotic BH-NS binaries which can open a new way in searching the missing BH-NS binaries in our Galaxy. By considering the possible dark matter density spikes formed around the primordial black holes in the BH-NS binaries, we show that extremely high temperature ($\sim 10^6$ K) could be maintained on the surface of the neutron stars due to effective dark matter capture. This interesting feature can also help reveal the nature of dark matter and possibly further improve the upper limit of dark matter scattering cross section well below $10^{-47}$ cm$^2$.  
\end{abstract}


\section{Introduction}
A recent analysis using the population synthesis model suggests that there are $\sim 10^5$ black hole-neutron star (BH-NS) binaries inside our Galaxy \citep{Olejak}. While some of the BH-NS binaries would merge to form black holes, many of these binaries remain and hide in our Galaxy. These binaries are regarded as a holy grail of astronomy because they enable us to test the validity of General Relativity and put constraints on alternative theories of gravity \citep{Stairs,Faucher}. We can also understand the gravitational wave emission due to these binaries \citep{Abbott2,Tsutsui} and the merging process of compact objects \citep{Kyutoku,Fakhry}. However, after searching these BH-NS binaries for decades, none of these binaries in our Galaxy has been found to date. We only discovered two events of the final merging process of extragalactic BH-NS binaries via gravitational wave detection in 2020 (GW200105 and GW200115) \citep{Abbott}.

On the other hand, we expect that dark matter particles would be scattered by large celestial objects. For example, previous studies have obtained some modest limits on dark matter scattering cross section by using the sun \citep{Leane}, the moon \citep{Garani,Chan}, the Earth \citep{Mack,Bramante,Chan2}, Jupiter \citep{Leane2}, exoplanets \citep{Leane3}, brown dwarfs \citep{Acevedo}, and white dwarfs \citep{McCullough,Dasgupta} as targets. In particular, neutron stars are considered to be the best targets because they have a very high density of nucleons, which may be able to capture dark matter particles effectively. Many previous studies have examined the possible observational consequences of a neutron star capturing dark matter \citep{McDermott,Garani2,Bhattacharya,Su}. One of the simplest indicators is the surface temperature of the neutron stars \citep{Raj}. Energy would be transferred to neutron stars through dark matter capture so they would maintain an equilibrium temperature of $\sim 10^3$ K. However, detecting this `low temperature' is very challenging because the surface area of a neutron star is very small (radius $R \sim 10$ km). The amount of blackbody radiation emitted is too small, which is below the maximum sensitivity of current best telescopes \citep{Raj}. The lowest temperature upper limit of neutron stars observed is $33000$ K \citep{Guillot,Raj}, which is still an order of magnitude larger than the expected equilibrium temperature due to dark matter capture.

Fortunately, some earlier theoretical studies have suggested the formation of dark matter density spike around a black hole due to conservation of angular momentum and radial action \citep{Gondolo,Merritt,Gnedin,Nampalliwar}. Even for primordial black holes, dark matter density spikes can also be formed around them \citep{Eroshenko,Ireland}. Recent analyses using the theory of dynamical friction have shown evidence of the existence of dark matter density spikes surrounding the primary supermassive black hole in OJ 287 \citep{Chan3}, and the stellar-mass black holes in A0620-00 and XTE J1118+480 \citep{Chan4}. The existence of dark matter density spikes in these systems can satisfactorily explain the observed orbital period decay. Therefore, we can conceive that dark matter density spikes might also exist around some of the black holes inside BH-NS binaries. This provides a huge amount of dark matter particles for the neutron star to capture, which could transfer a large amount of energy to the neutron star to maintain an extremely high equilibrium temperature. In this article, we show that the extremely high equilibrium temperature would be easily observable by current technologies. This may help search the missing BH-NS binaries in our Galaxy and provide a much more stringent constraint on dark matter scattering cross section compared with that obtained by XENON1T experiment \citep{Xenon}.

\section{Heating and Cooling of Neutron Stars}
Dark matter can be captured by a neutron star. The strong gravity of a neutron star would accelerate dark matter particles towards its center. Since the number density of a neutron star is very high $\sim 10^{45}$ cm$^{-3}$, dark matter particles would interact with the nucleons and finally lose most of their energy. After that, the dark matter particles would be gravitationally bounded inside the neutron star and eventually thermalized \citep{Bhattacharya}. A small dark matter halo would be accumulated inside the neutron star core \citep{Garani3,Bell}. In the followings, we will take the fiducial values of neutron star mass $M_{\rm NS}=1.3M_{\odot}$ and radius $R=11.6$ km.

Generally speaking, dark matter scattering inside a neutron star could be optically thick or optically thin, depending on the scattering cross section $\sigma_{\chi N}$. For $\sigma_{\chi N}> \sigma_{\rm crit}=2.5 \times 10^{45}$ cm$^2$, the scattering would be optically thick so that the capture rate would approach to the geometric capture rate \citep{Garani2}
\begin{eqnarray}
C_{\rm NS}&=&4.8 \times 10^{25}~{\rm s}^{-1} \left(\frac{\rho_{\chi}}{\rm GeV/cm^3} \right) \nonumber\\
&& \times \left(\frac{1~\rm GeV}{m_{\chi}} \right) \left( \frac{R}{11.6~\rm km} \right) \left(\frac{M_{\rm NS}}{1.3M_{\odot}} \right),
\end{eqnarray}
where $\rho_{\chi}$ is the dark matter density and $m_{\chi}$ is the mass of the dark matter particles. For $\sigma_{\chi N} \le \sigma_{\rm crit}$, the scattering would be optically thin and the capture rate becomes \citep{Bhattacharya}
\begin{eqnarray}
C_{\rm NS}&=&3.3\times 10^{25}~{\rm s^{-1}} \left(\frac{\rho_{\chi}}{\rm GeV/cm^3} \right) \left(\frac{1~\rm GeV}{m_{\chi}} \right) \nonumber\\ 
&& \times \left(\frac{\sigma_{\chi N}}{10^{-45}~\rm cm^2} \right) \left(\frac{v_{\rm esc}}{1.9\times 10^5~\rm km/s} \right)^2 \left(\frac{220~\rm km/s}{v_{\rm gal}} \right) \nonumber\\
&& \times \left(1-\frac{1-e^{-A^2}}{A^2} \right) \left(\frac{M_{\rm NS}}{1.3M_{\odot}} \right),
\end{eqnarray}
where $v_{\rm esc}$ is the escape velocity of the neutron star, $v_{\rm gal}$ is the galactic rotation velocity, and
\begin{equation}
A^2=\frac{6v_{\rm esc}^2m_{\chi}m_N}{v_{\rm gal}^2(m_{\chi}-m_N)^2},
\end{equation}
with $m_N$ being the nucleon mass.

After multiple scattering between the dark matter particles and nucleons, most of the kinetic energy of captured dark matter would be dissipated and transferred to the neutron star. Therefore, the rate of the energy dissipation is \citep{Bell}
\begin{equation}
\dot{E}_{\rm kin}=m_{\chi}c^2 \left(\frac{1}{\sqrt{B(0)}}-1 \right)C_{\rm NS},
\end{equation}
where $B(0)$ is the relativistic factor of the core region. If dark matter could self-annihilate, some of the captured dark matter would further contribute the rest mass energy to the neutron star. Therefore, the rate of the total energy transferred to the neutron star is $\dot{E}=\dot{E}_{\rm kin}+2 \Gamma_{\rm ann}m_{\chi}c^2$ \citep{Bell}. In equilibrium, the annihilation rate $\Gamma_{\rm ann}=C_{\rm NS}/2$. Therefore, including self-annihilation, the total energy transfer rate is \citep{Bell}:
\begin{equation}
\dot{E}=\frac{m_{\chi}c^2}{\sqrt{B(0)}}C_{\rm NS}.
\end{equation}

Normally, the major cooling of a neutron star is via blackbody radiation and neutrino cooling \citep{Bramante2}. Nevertheless, radiative cooling would dominate at a later stage. By neglecting the neutrino cooling, we can estimate the cooling time scale of a neutron star via the following equation \citep{Potekhin}: 
\begin{equation}
\dot{E}-4\pi R^2 \sigma_{\rm SB}T_{\rm eff}^4= \frac{\pi^2}{2} \left(\frac{M_{\rm NS}}{m_N} \right) \left(\frac{k^2T_b}{\epsilon_F}\right) \frac{dT_b}{dt},
\end{equation}
where $\sigma_{\rm SB}$ is the Stefan-Boltzmann constant, $T_{\rm eff}$ is the effective surface temperature of the neutron star, $T_b$ is the temperature of the inner boundary, and $\epsilon_F$ is the Fermi energy. The relation between the effective surface temperature and the inner boundary temperature is model-dependent. If we take the typical `fully-accreted' scenario, the relation can be given by \citep{Potekhin2}
\begin{equation}
\left(\frac{T_{\rm eff}}{10^6~\rm K}\right)^4=g_{14} \left(18.1 \frac{T_b}{10^9~\rm K} \right)^{2.42},
\end{equation}
where $g_{14}$ is the surface gravitational acceleration in the unit of $10^{14}$ cm/s$^2$. Due to the relativistic effect, the redshifted surface temperature observed is given by \citep{Bell}
\begin{equation}
T_{\infty}=\sqrt{B(R)}T_{\rm eff}=\sqrt{1-\frac{2GM_{\rm NS}}{Rc^2}}T_{\rm eff}.
\end{equation}
Therefore, by solving Eq.~(6), we can get the cooling curve of a neutron star if the initial temperature is set as $T_b=10^{10}$ K (see Fig.~1), assuming in the optically thick regime. We can see that after $\sim 50000$ yr, the surface temperature would achieve an equilibrium temperature due to dark matter heating. By taking $dT_b/dt=0$ in Eq.~(6), we can get this equilibrium redshifted temperature as \citep{Bell}
\begin{equation}
T_{\infty}=\left[\frac{B^2(R)\dot{E}}{4\pi \sigma_{\rm SB}R^2} \right]^{1/4}.
\end{equation}

\section{Enhanced dark matter capture in BH-NS binaries}
A recent study suggests evidence that the black hole binaries A0620-00 and XTE J1118+480 likely have dark matter density spikes \citep{Chan4}. At the positions of the companion stars, the dark matter density can be as high as $10^{-13}-10^{-11}$ g cm$^{-3}$ (i.e. $5\times 10^{10}-5\times 10^{12}$ GeV/cm$^3$) \citep{Chan4}. The dynamical friction due to the dark matter density spikes can satisfactorily explain the observed abnormal orbital period decay of the companion stars. Nevertheless, based on the benchmark model, dark matter density spikes are expected to form around supermassive black holes at galactic centers only. Stellar-mass black holes of an astrophysical origin are not able to form dark matter density spikes due to their much weaker gravitational influence \citep{Ireland}. Therefore, \citet{Ireland} has explored the possibility of a primordial origin for the stellar-mass black holes in A0620-00 and XTE J1118+480 to explain the existence of the dark matter density spikes. If these black holes are primordial black holes originally, dark matter can be effectively captured by the primordial black holes during the radiation dominated era to form dark matter density spikes \citep{Adamek,Carr}. These dark matter density spikes formed could be maintained for a long time unless there are astrophysical processess modifying the density distribution, such as tidal stripping and mergers \citep{Ireland}. Coincidentally, there is evidence that XTE J1118+480 might be formed through dynamical capture \citep{Gonzalez,Ireland}, which suggests the possibility that the black hole in XTE J1118+480 is indeed a primordial black hole and explains why it contains a dark matter density spike. 

Based on the above argument, we expect that some stars (including neutron stars and white dwarfs) could be captured by the primordial black holes with dark matter density spikes to form binaries. Hence, it can be conceived that dark matter density spikes would also exist in some of the BH-NS binaries. The extremely high dark matter density in the density spike structure would greatly enhance the dark matter capture rate and heat up the neutron star in a BH-NS binary to a very high temperature. The density spike model can be described by \citep{Lacroix,Kavanagh,Chan4}
\begin{equation}
\rho_{\chi}=\left\{
\begin{array}{ll}
0 & {\rm for }\,\,\, r\le 2R_s \\
\rho_0 \left(\frac{r}{r_{\rm sp}} \right)^{-\gamma} & {\rm for }\,\,\, 2R_s <r \le r_{\rm sp}, \\
\rho_0 & {\rm for}\,\,\, r>r_{\rm sp} \\
\end{array}
\right.
\end{equation}
where $R_s=2GM_{\rm BH}/c^2$ with $M_{\rm BH}$ being the black hole mass, $\gamma$ is the spike index ranging from $1.5-2.5$ \citep{Gondolo}, $\rho_0 \approx 0.4$ GeV/cm$^3$ is the local dark matter density, and $r_{\rm sp}$ is the spike radius given by \citep{Kavanagh}
\begin{equation}
r_{\rm sp}=\left[\frac{(3-\gamma)(0.2^{3-\gamma})M_{\rm BH}}{2\pi \rho_0} \right]^{1/3}.
\end{equation}

A recent analysis has shown that there are about $2.3\times 10^5$ BH-NS binaries in our Galactic disk \citep{Olejak}, although we have not detected any of them yet. The average masses of the black hole and neutron stars are $M_{\rm BH}=(12.8-15.9)M_{\odot}$ and $M_{\rm NS}=1.3M_{\odot}$ respectively. Using these values, the timescale for a BH-NS binary merger due to gravitational wave emission is $\sim 4$ Gyr \citep{Faucher}. Therefore, due to this large timescale, nearly 90\% of the BH-NS binaries would still remain in our Galaxy \citep{Olejak}. 

Suppose a considerable portion of the black holes in BH-NS binaries is primordial in origin, surrounding with dark matter density spikes. We take the benchmark values $v_{\rm gal}=220$ km/s, $\gamma=1.75$, and $M_{\rm BH}=12.8M_{\odot}$ and considering the neutron star is orbiting the black hole in a BH-NS binary with orbital radius $r \approx 0.1$ AU. Using Eq.~(10), we get $\rho_{\chi}=5.7\times 10^{-13}$ g cm$^{-3}$ (or $3.2\times 10^{11}$ GeV/cm$^3$). In the optically thick regime, using Eq.~(4), Eq.~(5) and Eq.~(9), the equilibrium redshifted surface temperature is $T_{\infty}=2.0 \times 10^6$ K for annihilating dark matter and $T_{\infty}=1.5 \times 10^6$ K for non-annihilating dark matter. In the optically thin regime, the equilibrium surface temperature depends on the scattering cross section (see Fig.~2). Therefore, if we can find out any BH-NS binary in our Galaxy and able to detect the surface temperature of the neutron star, we could constrain the scattering cross section. Moreover, the constrained scattering cross section almost does not depend on the dark matter mass for $m_{\chi}=1$ MeV to 100 TeV. This can give a very stringent upper limit for the scattering cross section, even much tighter than the current XENON1T upper limit \citep{Xenon} (see Fig.~3).

On the other hand, the redshifted surface temperature also depends on the dark matter density. Roughly speaking, if the neutron star is located closer to the black hole in a BH-NS binary, the dark matter capture rate would be higher due to a higher dark matter density. In Fig.~4, we plot the relation between $T_{\infty}$ against $\rho_{\chi}$ in the optically thick regime. If the dark matter density is $10^{-11}$ g cm$^{-3}$, $T_{\infty}$ can be as high as $4 \times 10^6$ K so the neutron star appears as an `ultra-violet' star. In fact, for XTE J1118+480, the dark matter density at the position of the companion star is $\rho_{\chi}=1.60^{+1.51}_{-0.73} \times 10^{-11}$ g cm$^{-3}$ \citep{Chan4}. Therefore, detecting this kind of ultra-violet star is quite possible.

\section{Detecting the exotic BH-NS binaries}
Although the expected redshifted surface temperature of the exotic neutron stars is very high $\sim 10^6$ K, their radii are too small so that the luminosity flux for each exotic neutron star may be too weak to observe. For $T_{\infty} \sim 10^6$ K, the peak frequency of radiation would be close to ultra-violet and X-ray bands. We first consider the XMM-Newton X-ray telescope. Its sensitivity limit is $\sim 10^{-14}$ erg cm$^{-2}$ s$^{-1}$ for the energy range $0.15-12$ keV (with exposure time $10^4$ s) \citep{Balaji}. It can detect any exotic BH-NS binaries with temperature larger than $10^5$ K within $\sim 20$ kpc (see Fig.~5). 

We also consider the Swift Ultra-Violet/Optical telescope (UVOT). The maximum detectable sensitivity is above magnitude $m_B=24.0$ (with 1000 s exposure time) \citep{Roming}. Based on this maximum sensitivity, we can determine the lowest temperature that the Swift UVOT can detect. For $T_{\infty} \sim 10^6$ K with the observing frequency $\nu=6.74\times 10^{14}$ Hz (bandwidth $=1.52 \times 10^{14}$ Hz), the Swift UVOT can only identify any exotic BH-NS binaries within 100 pc (see Fig.~5). 

Furthermore, we consider the current best telescope James Webb Space Telescope (JWST). The best sensitivity can allow the minimum flux density $\sim 0.5$ nJy (exposure time = $10^4$ s and signal-to-noise ratio = 2) \citep{Raj}. The lowest temperature that the JWST can identify is smaller compared to the Swift UVOT. For $T_{\infty}=10^6$ K with the observing frequency $\nu=1.5\times 10^{14}$ Hz (bandwidth $=3.5\times 10^{13}$ Hz), it can identify any exotic BH-NS binaries within 1 kpc (see Fig.~5). Note that any possible background and foreground noise has not been taken into account in the above estimations. In particular, the gas column density might contribute 10\% absorption for the infrared flux observed \citep{Raj}. Therefore, our estimations would be somewhat optimistic. 

Assume that our Galactic disk contains $\sim 10^5$ BH-NS binaries with dark matter density spikes. On average there is one exotic BH-NS binary within the radius of $\sim 130$ pc, assuming the local stellar density $=0.1$ pc$^{-3}$. Therefore, a considerable amount of exotic BH-NS binaries could be observed by the the XMM-Newton telescope and JWST within the radius of 1 kpc. We predict that this method can effectively find out the missing BH-NS binaries in our Galaxy and help constrain the parameters of dark matter. 

Note that there are some known astrophysical mechanisms of neutron star heating which can also generate high temperature, such as vortex creep heating \citep{Fujiwara}. Therefore, it is not sufficient to claim the discovery of any BH-NS binaries and constrain dark matter parameters when we have confirmed any neutron stars with $T_{\infty}\sim 10^6$ K. Nevertheless, finding out very hot neutron stars would be one of the important features showing the possibility of dark matter capture in neutron stars. If we can identify any very hot neutron star candidates, we can use other methods (e.g. data of Shapiro delay) to confirm whether the companions are black holes or not. If they are black holes, we can check whether dark matter density spikes exist by examining the effect of dynamical friction \citep{Chan4}. Therefore, identifying very hot neutron stars is the first crucial step to find out galactic BH-NS binaries. Moreover, for the BH-NS binary consisting of a $13M_{\odot}$ black hole and $1.3M_{\odot}$ neutron star companion with semi-major axis $a \sim 0.1$ AU, the time period would be about 3 days. If the orbital plane is along the line of sight, this can potentially lead to periodic signal amplification in the luminosity observed (infrared or X-ray) due to the self-lensing effect \citep{Kruse,Chan5}. The amplification would be the largest when the neutron star is inside the black hole's Einstein ring region (with radius $R_E=\sqrt{4GM_{\rm BH}a/c^2}$) along the line of sight. The maximum amplification factor is $A_L=\sqrt{1+4R_E^2/R^2} \approx 5860$ \citep{Rahvar}. Examining time series data could shed light for such configurations and help search the missing BH-NS binaries. However, this feature can only be seen if the orbital plane is very close to an edge-on configuration ($a \cos i \le R_E$, where $i$ is the orbital inclination angle), which requires $89.87^{\circ} \le i \le 90.13^{\circ}$. For random distribution of $i$ in BH-NS binaries, the probability of finding this special configuration would be $0.0014$ only.

\section{Discussion}
In this article, by using the idea of the dark matter density spike around black holes, we expect that the equilibrium redshifted surface temperature of the neutron stars inside the exotic BH-NS binaries can be as high as $10^6$ K. This temperature can be detected by the JWST within the distance of 1 kpc with affordable exposure time. Therefore, searching for extremely high temperature of neutron stars might be a new way to search for the missing BH-NS binaries, which is complementary to the traditional search using radio telescopes. If we can find out any exotic BH-NS binary using this method, we can simultaneously improve our understanding about compact objects, merging process, gravitational wave physics, and constrain dark matter properties. It can also help understand the model of dark matter admixed neutron star proposed recently \citep{Kain}. Therefore, this can open up many new lines of study and new windows of observations.

As the cooling timescale is short enough $\sim 10^{4}$ yr, we expect that most of the neutron stars in BH-NS binaries containing dark matter spikes should have achieved equilibrium redshifted surface temperature of $10^5-10^6$ K, unless the dark matter scattering cross section is much smaller than $10^{-47}$ cm$^2$. If there is no enhanced dark matter capture due to dark matter density spike, the equilibrium surface temperature should be $\sim 2000$ K or below (assuming the local dark matter density $\rho_{\chi}=0.4$ GeV/cm$^3$) \citep{Bramante2}. However, detecting this low temperature based on current technology is very challenging. 

Moreover, in the optically thin regime, one can see that we cannot directly constrain dark matter cross section based on the redshifted surface temperature only. The relation also depends on the dark matter density. Nevertheless, if we can identify any potential BH-NS binary, we may constrain the dark matter density by observing the orbital period decay, like the analyses done in \citet{Chan4,Chan3}. This can help control the dark matter density factor to constrain dark matter scattering cross section based on the redshifted surface temperature. 

In fact, current observed coldest neutron star, PSR J2144-3933, has $T_{\infty} \le 33000$ K and is located at $d=172^{+20}_{-15}$ pc from us \citep{Guillot,Raj}. It is an isolated neutron star without enhanced dark matter capture. We anticipate that observing farther away to the distance 1 kpc from us can possibly figure out the first exotic BH-NS binary to verify our proposal and create a new page for understanding dark matter physics and relativistic compact systems. This new method can simultaneously kill two birds (finding the first BH-NS binary and constraining dark matter) with one stone (searching for the compact objects with extremely high temperature).

\begin{figure}
\vskip 5mm
 \includegraphics[width=140mm]{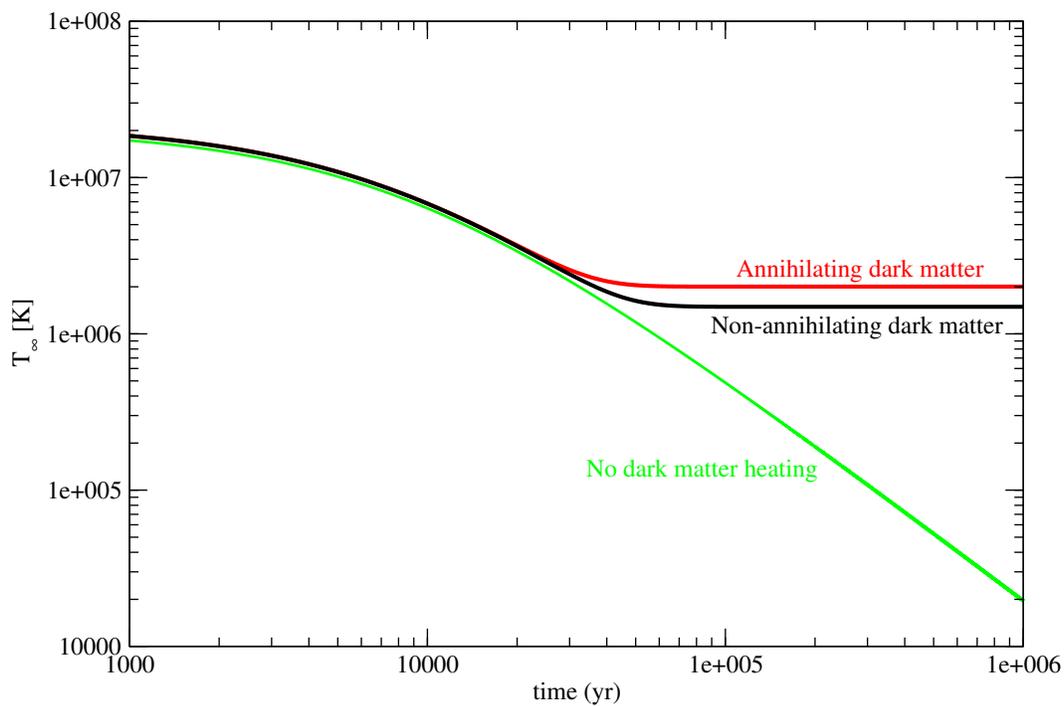} 
\caption{The redshifted surface temperature $T_{\infty}$ against time in the optically thick regime. The red, black, and green solid lines respectively represent the cooling curves involving annihilating dark matter heating, non-annihilating dark matter heating, and without dark matter heating. Here, we have assumed $\rho_{\chi}=5.7 \times 10^{-13}$ g cm$^{-3}$.}
\label{Fig1}
\end{figure}

\begin{figure}
\vskip 5mm
 \includegraphics[width=140mm]{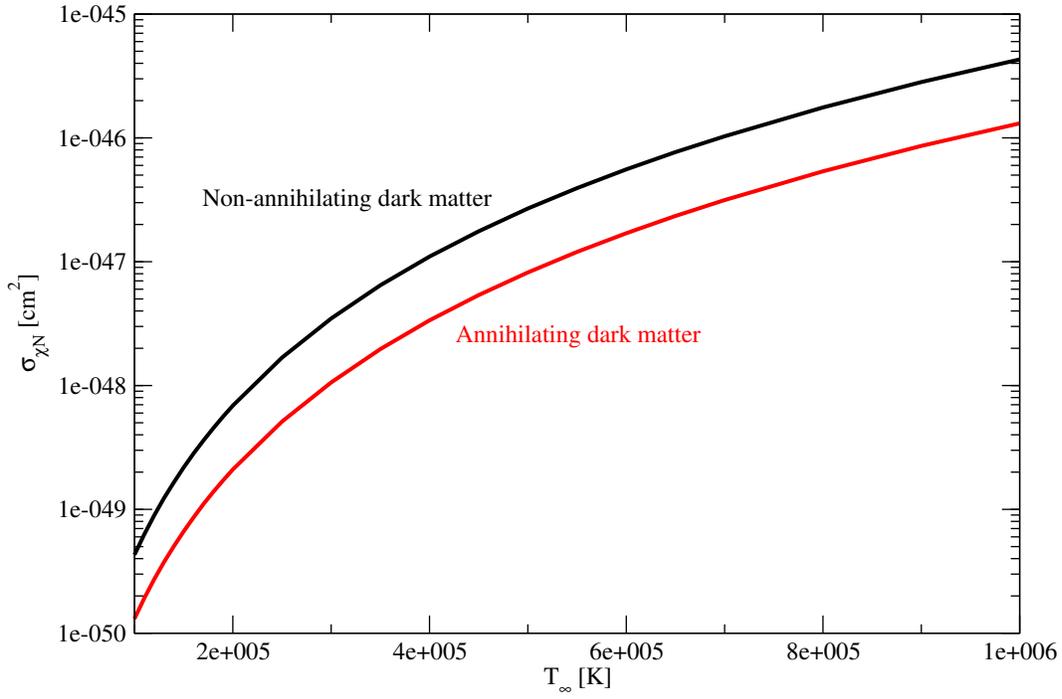}
\caption{The upper limits of $\sigma_{\chi N}$ for annihilating dark matter heating scenario (red) and non-annihilating dark matter heating scenario (black) in the optically thin regime. Here, we have assumed $\rho_{\chi}=5.7 \times 10^{-13}$ g cm$^{-3}$.}
\label{Fig2}
\end{figure}

\begin{figure}
\vskip 5mm
 \includegraphics[width=150mm]{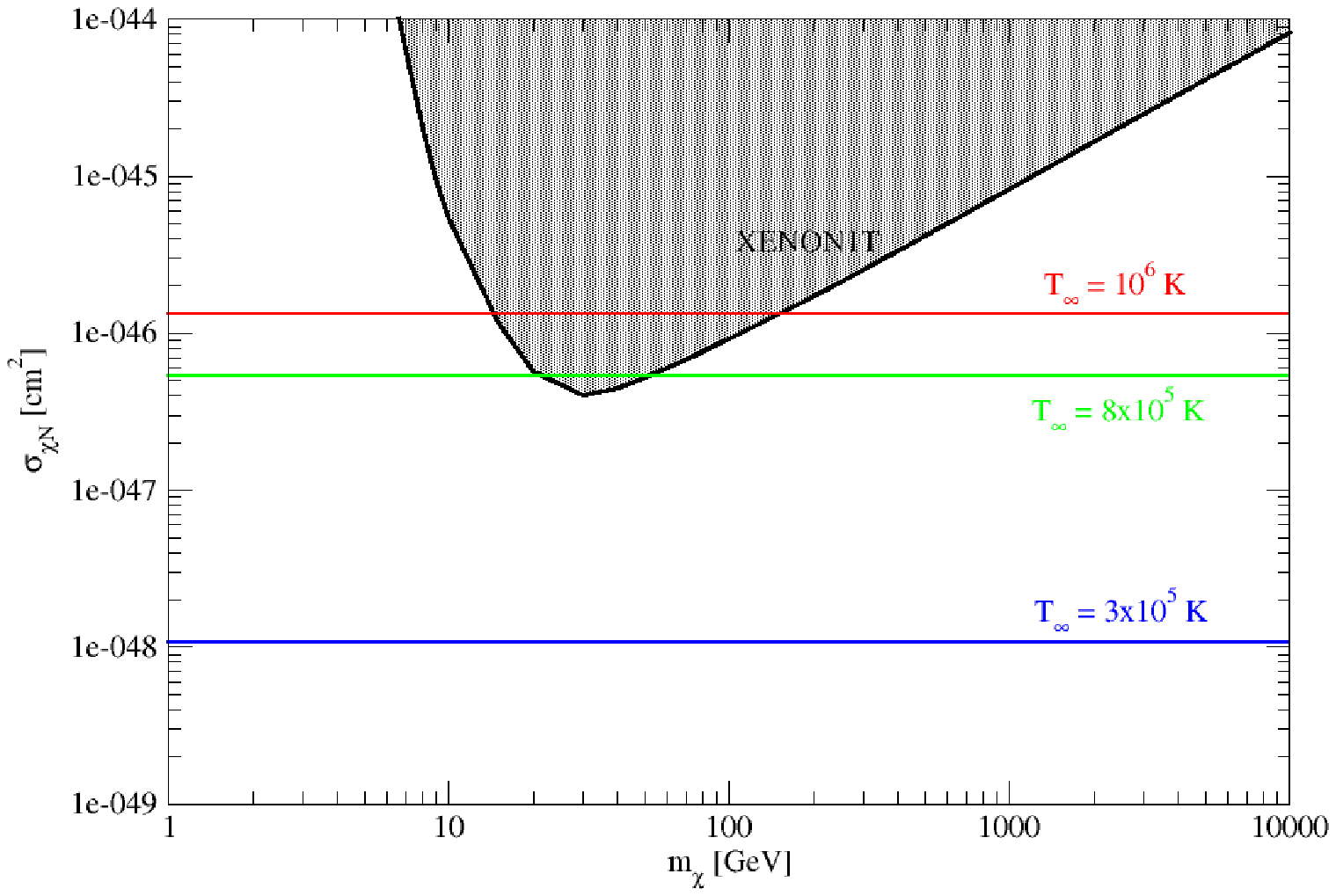}
\caption{The shaded area bounded by the black line indicates the forbidden parameter space of $\sigma_{\chi N}-m_{\chi}$ (spin-independent) obtained from the XENON1T experiment \citep{Xenon}. The red, green, and blue solid lines represent the upper limits of $\sigma_{\chi N}$ for the equilibrium redshifted surface temperature $T_{\infty}=10^6$ K, $8\times 10^5$ K and $3\times 10^5$ K respectively in the optically thin regime (heated by annihilating dark matter). Here, we have assumed $\rho_{\chi}=5.7 \times 10^{-13}$ g cm$^{-3}$.}
\label{Fig3}
\end{figure}

\begin{figure}
\vskip 5mm
 \includegraphics[width=140mm]{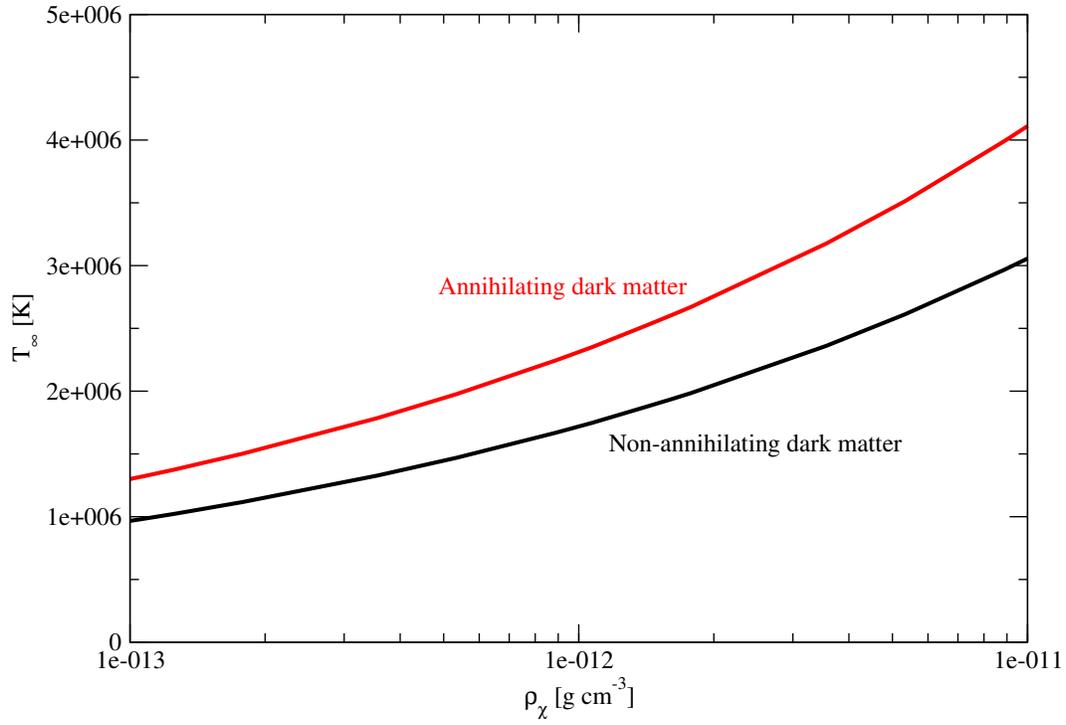} 
\caption{$T_{\infty}$ against $\rho_{\chi}$ in the optically thick regime.}
\label{Fig4}
\end{figure}

\begin{figure}
\vskip 5mm
 \includegraphics[width=140mm]{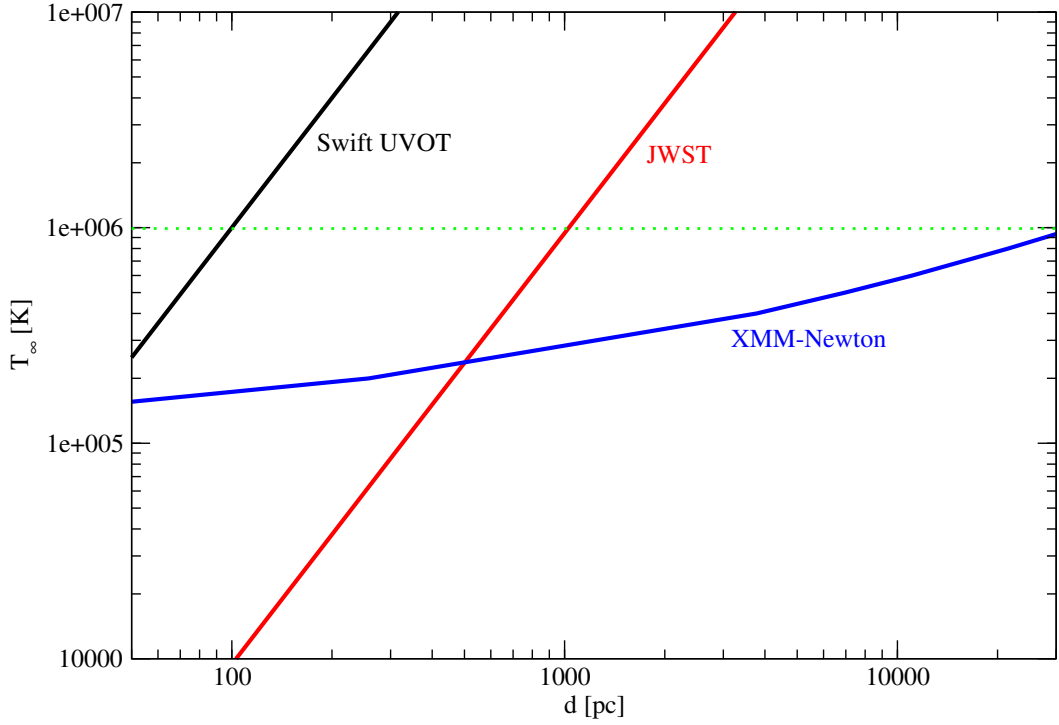} 
\caption{The minimum $T_{\infty}$ that can be observed by the Swift UVOT (black line), XMM-Newton (blue line) and JWST (red line) against the distance $d$ from us. The green dotted line indicates the order of magnitude of $T_{\infty}$ predicted. Here, the observing frequency ranges of the Swift UVOT, XMM-Newton and JWST are $(5.22-8.26) \times 10^{14}$ Hz, $(3.63-283) \times 10^{16}$ Hz and $(1.15-1.85) \times 10^{14}$ Hz respectively.}
\label{Fig5}
\end{figure}

\section{Acknowledgements}
We thank the anonymous referee for the useful comments and suggestions, including the discussion of the time series data examination. The work described in this paper was partially supported by the Dean's Research Fund of The Education University of Hong Kong and the grants from the Research Grants Council of the Hong Kong Special Administrative Region, China (Project No. EdUHK 18300922 and EdUHK 18300324).

\end{document}